# Disordered Fe vacancies and superconductivity in potassium-intercalated iron selenide (K$_{2-x}$Fe$_{4+y}$Se$_5$)

Chih-Han Wang[1,2*], Ta-Kun Chen[1*], Chung-Chieh Chang[1], Chia-Hao Hsu[1], Yung-Chi Lee[1,3], Ming-Jye Wang[4], Phillip M. Wu[5] and Maw-Kuen Wu[1,3,6]†

**Affiliations:**

[1] Institute of Physics, Academia Sinica, Taipei, Taiwan
[2] Department of Electronic and Computer Engineering, National Taiwan University of Science and Technology, Taipei, Taiwan
[3] Department of Physics, National Tsing Hua University, Hsinchu, Taiwan
[4] Institute of Astrophysics and Astronomy, Academia Sinica, Taipei, Taiwan
[5] Deparment of Applied Physics and Geballe Laboratory for Advanced Materials, Stanford University, Stanford, CA, USA
[6] National Donghwa University, Hualien, Taiwan
*These authors contributed equally to this work.
† Corresponding author. **E-mail:** mkwu@phys.sinica.edu.tw

**Abstract:**
The parent compound of an unconventional superconductor must contain unusual correlated electronic and magnetic properties of its own. In the high-Tc potassium intercalated FeSe, there has been significant debate regarding what the exact parent compound is. Our studies unambiguously show that the Fe-vacancy ordered K$_2$Fe$_4$Se$_5$ is the magnetic, Mott insulating parent compound of the superconducting state. Non-superconducting K$_2$Fe$_4$Se$_5$ becomes a superconductor after high temperature annealing, and the overall picture indicates that superconductivity in K$_{2-x}$Fe$_{4+y}$Se$_5$ originates from the Fe-vacancy order to disorder transition. Thus, the long pending question whether magnetic and superconducting state are competing or cooperating for cuprate superconductors may also apply to the Fe-chalcogenide superconductors. It is believed that the iron selenides and related compounds will provide essential information to understand the origin of superconductivity in the iron-based superconductors, and possibly to the superconducting cuprates.

**One Sentence Summary:**

Our studies unambiguously show that the Fe-vacancy ordered K$_2$Fe$_4$Se$_5$ is the magnetic, Mott insulating parent compound of the superconducting state, and non-superconducting K$_2$Fe$_4$Se$_5$ becomes a superconductor after high temperature annealing, strongly indicating that superconductivity in K$_{2-x}$Fe$_{4+y}$Se$_5$ originates from the Fe-vacancy order to disorder transition.

**Main Text:**

The superconducting transition temperature, Tc, of the Fe chalcogenide superconductors, first discovered in 2008 (1), now rivals those of some cuprates. Many new exciting results (2-5) have transpired since the initial report in a relatively short amount of time, indicating the extreme interest in this system. Tc was enhanced dramatically by applying pressure (6), and then stabilized in bulk with Tc up to 46 K after alkali-intercalation between the iron selenide layers (A$_{1-x}$Fe$_{2-y}$Se$_2$) (7-10). The high-Tc in this system is accompanied by an intrinsic multiphase nature, with the presence of iron vacancy order in the non-superconducting regime (11-14). More recently, the observation of superconducting energy gap of ~20 meV in monolayer FeSe on $SrTiO_3$ (STO) substrate (15) puts Tc of Fe-chalcogenide higher than other Fe-based superconductors. Despite extensive studies, many critical issues remain unresolved.

It is generally accepted that there exists many competing states, such as orbital order, magnetic order, nematic electronic phases, structural distortion and superconductivity in the Fe-based superconductors. However, which interaction is at the root of observed phenomena and superconductivity? The large anisotropy of properties found below the structural phase transition experimentally (16-19) suggests the structural phase transition is driven by electron correlation (20, 21), and the corresponding fluctuations might also be related to the pairing mechanism of superconductivity (22, 23). One candidate responsible for the structural phase transition is the electronic nematicity (24-26). Another candidate as the underlying common mechanism for the structural and antiferromagnetic transitions in the Fe-based superconductors is the orbital ordering (27).

Another issue is the controversy of the superconducting phase in $A_{1-x}Fe_{2-y}Se_2$ system. Apart from the stable $A_2Fe_4Se_5$ non-superconducting matrix in superconducting crystals, a separated phase with an expanded $c$-axis and a composition close to $A_xFe_2Se_2$ (28-34) was assigned to be the superconducting phase with $T_C$ ~ 29–32 K. Others attribute the superconductivity to originate from a parent phase of semiconducting antiferromagnetic $A_2Fe_3Se_4$ with rhombus ($\sqrt{2} \times 2\sqrt{2}$) Fe-vacancy order (35), of $A_2Fe_7Se_8$ parallelogram structure (36), or of $A_3Fe_4Se_6$ instead of the insulating antiferromagnetic $A_2Fe_4Se_5$ with $\sqrt{5} \times \sqrt{5}$ Fe-vacancy order. Pinpointing the parent compound will help to address the issue of competing states and mechanisms for superconductivity.

ARPES investigations on alkaline-metal iron selenides ($A_xFe_{2-y}Se_2$) reveal the absence of hole pockets at the Fermi surface (FS) (37, 38). This striking result challenges the widely believed FS nesting picture. In their ARPES study on $A_xFe_{2-y}Se_2$ (A = K, Rb), Yi et al. (39) concluded that their results could be consistently understood as a temperature-induced crossover from a metallic state at low temperatures to an orbital-selective Mott phase at high temperatures. They further pointed out that the superconducting state of $A_xFe_{2-y}Se_2$ is near the boundary of such an orbital-selective Mott phase constrains the system to have sufficiently strong on-site Coulomb interactions and Hund's coupling, highlighting the nontrivial role of electron correlation in this family of iron-based superconductors. Zhang *et al.* showed that the presence of abundant Raman modes in $A_xFe_{2-y}Se_2$ (40) is the consequence of the superstructure with ordered iron vacancies. They further identified in $A_xFe_{2-y}Se_2$ a broad, asymmetric peak around 1600 $cm^{-1}$ as process involving optical magnons (41). The intensity of this two-magnon peak falls sharply on entering the superconducting phase suggesting a complete mutual proximity effect occurring within a microscopic structure based on nanoscopic phase separation.

A recent study on the non-superconducting $Fe_4Se_5$ (36), which exhibits the $\sqrt{5} \times \sqrt{5}$ Fe-vacancy order and is magnetic, shows that it becomes superconducting after high temperature annealing. This observation strongly indicates that the long pending question whether magnetic and superconducting state are competing or cooperating may also apply to the Fe-chalcogenide superconductors. Consequently, the question whether the Fe-vacancy order to disorder transition can be associated with the presence of superconductivity becomes essential.

Addressing this question in pure FeSe may be complicated by issues of different stoichiometry. Here we report a systematic study of $K_{2-x}Fe_{4+y}Se_5$ compounds with controlled stoichiometry, and we unambiguously demonstrate the appearance and disappearance of superconductivity in the same sample depending on preparation processes. Our measurements clearly show the correlation between the Fe-vacancy disorder and superconductivity in these compounds. Furthermore, we pinpoint the exact phase that is the parent compound from which superconductivity can emerge. These results along with recent theories provide a framework for better understanding the origin of superconductivity in Fe-chalcogenides.

A novel synthesis approach developed in this report is to combine the mechanical ball milling and high temperature calcination processes to the formation of polycrystalline $K_xFe_{2-y}Se_2$ samples with homogenous chemical composition and crystal phase (Supplement Materials and Method). As we discuss below, our samples maintain the 245 stoichiometry, with no presence of other phases such as the 122 phase. We treated the sealed powder material with $K_2Fe_4Se_5$ stoichiometry (we denote as $2(Fe_4)5$) into furnace by slowly heating to 700 ºC in 7 hours, kept at the same temperature for 24 hours and then slowly cooled to room temperature. These as grown samples (-AG) were further heat treated with two processes. The first is a low temperature post-annealing at 300 ºC for 12 hours followed by ice water quench (-LT). The second is a high temperature (with furnace already at 750 ºC) treatment for 3-6 hours followed by quenching in ice water (-HT). In Fig. 1(a) we show the evolution of the temperature dependent magnetic susceptibility for $2(Fe_4)5$ samples, the powder X-ray diffraction (XRD) data are shown in Supplementary Fig. S1. The as-grown sample, $2(Fe_4)5$-AG, shows no feature below 300 K. After low temperature post-annealing at 300 ºC, the sample ($2(Fe_4)5$-LT) exhibits an anomalous transition at 125 K, which we discuss more in depth below. This transition could be completely eliminated by further high temperature heat-treatment at 750 ºC ($2(Fe_4)5$-HT) and quenched to ice water. The $2(Fe_4)5$-HT sample reveals no magnetic feature in between 50–300 K. Instead, two diamagnetic-like transitions were observed at 29 K and 11 K, suggesting the presence of minor superconducting phases. Importantly, the observed 125 K transition and the superconducting-like transitions can be switched back and forth in the same $2(Fe_4)5$ sample by cycling through the LT, HT heat-treatment process.

Since the stoichiometric $2(Fe_4)5$ sample has intrinsic 20% Fe-vacancy, it is difficult to completely suppress the $\sqrt{5} \times \sqrt{5}$ Fe-vacancy order, explaining the low superconducting volume fraction observed. We introduce extra iron and intentionally reduce the potassium in an attempt to fill the vacant sites and allow additional iron to go into the crystal structure. The nominal compositions for these samples are $K_2Fe_{4.1}Se_5$ ($2(Fe_{4.1})5$), $K_{1.9}Fe_{4.1}Se_5$ ($1.9(Fe_{4.1})5$), $K_{1.9}Fe_{4.2}Se_5$ ($1.9(Fe_{4.2})5$) and $K_{1.9}Fe_{4.3}Se_5$ ($1.9(Fe_{4.3})5$).

Similar to $2(Fe_4)5$, the other samples can be switched from the state with 125 K transition to superconducting by the heat treatment cycle, as shown for $2(Fe_{4.1})5$ in Fig. 1(b). Supplementary Fig. S2 displays the comparison of the magnetic behavior of $2(Fe_4)5$, $1.9(Fe_{4.1})5$ and $1.9(Fe_{4.2})5$ high temperature annealed then ice water quenched samples. The superconducting volume increases with reduced potassium and increased iron, with bulk superconductivity emerging in the $K_{1.9}Fe_{4.2}Se_5$ as grown samples after keeping it at 750 ºC for several hours. We note that for the HT process, it is important to place samples in a pre-heated furnace. We also found that longer annealing time (>10 hours) resulted in poorer superconducting volume ratio, which is likely due to the degradation of the sample from the reaction between potassium and quartz tube. Additionally, extra iron such as $1.9(Fe_{4.3})5$ also does not increase the superconducting volume ratio, as shown in Supplementary Fig. S3.

We performed structural analyses with XRD and transmission electron microscopy (TEM) in order to understand the influence of the preparation process on the crystal structure and Fe-vacancy. Supplementary Fig. S4 shows the XRD patterns obtained from four $K_{2-x}Fe_{4+y}Se_5$ samples. The XRD pattern for $2(Fe_4)5$-LT is consistent with the result reported (42), with clear superstructure peaks fit to the $\sqrt{5} \times \sqrt{5}$ Fe-vacancy order and no superconductivity observed. When the nominal composition is altered to less potassium and more iron, a drastic reduction of the intensity of the superstructure peaks is observed in the inset of Fig. S4. For the $1.9(Fe_{4.2})5$-HT-6h sample, the superstructure peaks are no longer visible, and a superconducting transition is measured in susceptibility. The most important feature in our measurements is that there is no second phase from different stochiometry observable in all samples studied within the instrument resolution, unlike the previously reported superconducting crystals where phase separation is inevitable (34).

Based on the main Bragg peaks we calculated the lattice parameters using I4/m symmetry for samples with superstructure peaks (in $2(Fe_4)5$-LT and $1.9(Fe_{4.1})5$-HT samples) and I4/mmm symmetry for samples without superstructures ($1.9(Fe_{4.2})5$-HT). The results are summarized in Table 1 of Supplementary and Fig. S5. The data reveals clearly the reduction of lattice constant a (and b) and the elongation of lattice constant c in the superconducting samples. Our observations of the lattice parameter variation in the superconducting samples are consistent with those reported in phase-separated superconducting crystals, which apart from non-superconducting $2(Fe_4)5$ phase, contain extra phase with a compressed a-axis and an extended c-axis (12, 29).

We stress again that our crystals do not contain phases with different stoichiometry. Furthermore, the XRD patterns of the HT and LT phases for the same sample $1.9(Fe_{4.2})5$ are compared in Supplementary Fig. S6. The $\sqrt{5} \times \sqrt{5}$ superstructure peaks are present for the sample annealed at 300 °C, but are absent for the HT treatment, when bulk superconductivity emerges. Therefore, we believe that superconductivity is highly related to a situation without Fe-vacancy order, but not necessarily without Fe vacancies.

The relative intensity of (123) and (002) peaks is significantly changed from the non-superconducting to superconducting samples, as shown in Figs. S4 and S6. This can be understood if the Fe-4(d) vacant sites are gradually filled in the superconducting sample when the vacancies become more disordered so that the material becomes more two-dimensional. This picture is consistent with the SEM observation that the superconducting samples contain larger layer-type grains, whereas granular-type grains exist in the non-superconducting samples, see Supplementary Fig. S7. This grain morphology evolution certainly contributes to the peak intensity change between (123) and (002). Though a superconducting phase $K_yFe_2Se_2$ (y~0.5), which has been reported from Rietveld structural refinement and electron diffraction studies (29), can also account for the increase in (002)/(123) ratio with superconducting volume, we do not observe the presence of the $K_yFe_2Se_2$ phase in our sample with highest superconducting volume fraction.

Figure 2(a) shows a TEM image of the $2(Fe_4)5$-AG sample. Two vacancy orders were observed in this sample as shown in Fig. 2(b-d): a typical $\sqrt{5} \times \sqrt{5}$ Fe-vacancy order and a K-vacancy order that could be attributed to two possible origins (13). Although we do not determine which order, $2(Fe_4)5$-AG clearly has local inhomogeneity of composition resulting in the formation of K-vacancy order. Post annealing of the as-grown sample eliminates the extra K-vacancy order. Supplementary Fig. 8 shows two grains of a

2(Fe$_4$)5-LT sample exhibiting only superstructure reflections corresponding to the √5 × √5 Fe-vacancy order without any K-vacancy order.

In Fig. 2(h), the TEM for 1.9(Fe$_{4.2}$)5 is shown. The c-axis zone-pattern, easily obtained in TEM, corroborates the picture from SEM that the superconducting samples have more layered crystal morphology. The SAED pattern in Fig. 2(i) clearly shows no vacancy order. This is in very good agreement with the XRD analysis, that the Fe-vacancy order is successfully suppressed in samples with a stoichiometric composition close to the parent K$_2$Fe$_4$Se$_5$ phase. In other words, there is an order to disorder transition without a change in stoichiometry.

The features in the susceptibility as a function of temperature can be correlated to those in the resistivity versus temperature. The resistivity of the studied samples are shown in Fig. 3. Both 2(Fe$_4$)5-LT and 2(Fe$_4$)5-HT behave like a semiconductor, as shown in Fig. 3(a). We note that 2(Fe$_4$)5-HT shows superconducting-like transitions in susceptibility but with low volume fraction. The resistivity suggests a picture of extremely low density of disconnected superconducting puddles. The inset of Fig. 3(a) demonstrates that the temperature dependence of resistivity follows the Mott variable-range-hopping behavior, $\ln(\rho) \sim T^{(-1/3)}$. In 2(Fe$_4$)5-LT, the resistivity becomes extremely high, indicating most carriers are localized, at the temperature close to the transition observed in susceptibility in Fig. 1(a). Annealing the sample at high temperature and then quenching, 2(Fe$_4$)5-HT, reduces the sample resistivity. Extra iron that fills the vacancies also reduces the resistivity. For 1.9(Fe$_{4.1}$)5-LT sample that exhibits Fe-vacancy order, the temperature dependence of resistivity resembles data in the literature, showing a semiconducting to metallic behavior at intermediate temperature before the onset of superconducting transition at ~ 20 K (blue curve in Fig. 3(b)). The (1.9(Fe$_{4.1}$)5)-HT sample that has weaker superlattice peaks shows metallic behavior without a clear resistive peak at high temperature and becomes superconducting at an onset of 31 K (red curve in Fig. 3(b)). For the (1.9(Fe$_{4.2}$)5)-HT sample with highest superconducting volume ratio, its resistivity is essentially metallic, with temperature dependence similar to that observed in the FeSe superconductor, but with higher and sharp superconducting temperature at 31 K, the red curve in Fig. 3(c). Interestingly, the semiconducting-to-metallic behavior appears again after annealed at low temperature, (blue curve in Fig. 3(c)). Our results clearly demonstrate that the intrinsic resistivity of the superconducting sample of the K$_{2-x}$Fe$_{4+y}$Se$_5$ system is metallic without the resistive peak at high temperature. The observed resistive peak is due to the presence of a semiconducting phase that has ordered Fe-vacancy.

These results unambiguously resolve the controversy of the exact parent phase in K$_{2-x}$Fe$_{4+y}$Se$_5$ system. The K$_2$Fe$_4$Se$_5$ with Fe-vacancy order is the parent semiconducting phase for the Tc ~ 30 K superconducting phase. Superconductivity emerges after the removal of the vacancy order, which can be easily achieved by thermal processing. Our results also suggest the reduction in the Fe-vacancy density alone by adding extra irons into the lattice is not sufficient to induce bulk superconductivity. Introducing extra irons in the lattice makes the Fe-vacancy in the sample more disordered and leads to higher volume fraction of superconductivity. An important question that arises is what is the role of the interface between the disordered and ordered phases, and whether stabilizing the proper nanoscale interface is related to the very high Tc observed in monolayer FeSe.

This picture associating the superconducting transition with the Fe-vacancy order–disorder transition provides the basis to understand the results of the high-pressure study on the mixed phase $A_xFe_{2-y}Se_2$ (43). Sun et al. observed in their samples the suppression of superconductivity with increasing pressure. At high pressures, a second superconducting transition with higher Tc was induced. These observations can be understood in terms of the presence of two phases that have different Fe-vacancy order in the sample. Our results also support the viewpoint that $A_xFe_{2-y}Se_2$ tends towards mesoscopic phase separation into superconducting, semiconducting or AFM insulating phase proposed by Chen et al. (44) based on their $K_xFe_{2-y}Se_2$ ARPES study. It certainly will be essential to study in detail how the Fermi surface evolves from the non-superconducting parent state with Fe-vacancy order to the disordered superconducting state in the well-characterized $K_{2-x}Fe_{4+y}Se_5$ samples.

The mechanism behind the 125 K transition in samples with vacancy order is not well understood at the moment and requires further study. Preliminary magnetic susceptibility measurements indicate this transition is phenomenologically similar to that of the Verwey transition observed in $Fe_3O_4$, suggesting the possibility of charge/spin order in the non-superconducting parent compound. This is consistent with the electronic nematicity, which is associated with the C4 rotational symmetry breaking that has been observed in the Fe-based superconductors.

In summary, our studies on the $K_{2-x}Fe_{4+y}Se_5$ superconductors show that the magnetic, non-superconducting $K_2Fe_4Se_5$ phase with iron-vacancy order is a Mott insulator. Non-superconducting $K_2Fe_4Se_5$ becomes a superconductor after high temperature annealing, which disorders the Fe vacancy order but with no stoichiometry change. These results settle two issues: 1) that the Fe-vacancy ordered $K_{2-x}Fe_{4+y}Se_5$ is the insulating parent compound of the superconducting state and 2) superconductivity in $K_{2-x}Fe_{4+y}Se_5$ originates from the Fe-vacancy order to disorder transition. The similarities for this system to the cuprates are striking. Thus, the long pending question whether magnetic and superconducting state are competing or cooperating for cuprate superconductors may also apply to the Fe-chalcogenide superconductors. It is believed that the iron selenides and related compounds will provide essential information to understand the origin of superconductivity in the iron-based superconductors, and possibly to the superconducting cuprates.

References and Notes:
1. F. C. Hsu et al., Superconductivity in the PbO-type structure α-FeSe. *Proc. Natl. Acad. Sci. U.S.A.* **105**, 14262–14264 (2008).
2. M. K. Wu et al., The development of the superconducting PbO-type b-FeSe and related compounds. *Physica C* **469**, 340-349 (2009).
3. M. K. Wu, M. J. Wang, K. W. Yeh, Recent advances in β-FeSe$_{1−x}$ and related Superconductors. *Sci. and Technol. Adv. Mater.* **14**, 014402 (2013).
4. K. Deguchi, Y. Takano, Y. Mizuguchi, Physics and chemistry of layered chalcogenide superconductors. *Sci. Technol. Adv. Mater.* **13**, 054303 (2012).
5. E. Dagotto, The unexpected properties of alkali metal iron selenide superconductors. *Rev. Mod. Phys.* **85**, 849-867 (2013).
6. Y. Mizuguchi, et al., Superconductivity at 27K in tetragonal FeSe under high pressure. *Appl. Phys. Lett.* **93,** 152505 (2008).
7. J. Guo et al., Superconductivity in the iron selenide $K_xFe_2Se_2$ (0≤x≤1.0). *Phys. Rev. B* **82**, 180520 (2010).
8. Z. Wang et al., Microstructure and ordering of iron vacancies in the superconductor system $K_yFe_xSe_2$ as seen via transmission electron microscopy. *Phys. Rev. B* **83**, 140505(R) (2011).

9. A. Ricci *et al.*, Intrinsic phase separation in superconducting $K_{0.8}Fe_{1.6}Se_2$ ($T_c$ = 31.8 K) single crystals. *Supercond. Sci. Technol.* **24**, 082002 (2011).
10. T. P. Ying *et al.*, Observation of superconductivity at 30~46K in $A_xFe_2Se_2$ (A=Li, Na, Ba, Sr, Ca, Yb, and Eu). *Sci. Rep.* **2**, 426 (2012).
11. F. Ye *et al.*, Common Crystalline and Magnetic Structure of Superconducting $A_2Fe_4Se_5$ (A = K, Rb, Cs, Tl) Single Crystals Measured Using Neutron Diffraction. *Phys. Rev. Lett.* **107**, 137003 (2011).
12. W. Li *et al.*, *Phase separation and magnetic order in K-doped iron selenide superconductor. Nat. Phys.* **8**, 126-130 (2012).
13. Z. W. Wang *et al.*, Structural Phase Separation in $K_{0.8}Fe_{1.6+x}Se_2$ Superconductors. *J. Phys. Chem. C* **116**, 17847-17852 (2012).
14. Y. J. Yan *et al.*, Electronic and magnetic phase diagram in $K_xFe_{2-y}Se_2$ superconductors. *Sci. Rep.* **2**, 212 (2012).
15. Q.Y. Wang *et al.*, Interface-induced high-temperature superconductivity in single unit-cell FeSe films on SrTiO3. *Chin. Phys. Lett.* **29**, 037402 (2012).
16. J. H. Chu *et al.*, In-Plane Resistivity Anisotropy in an Underdoped Iron Arsenide Superconductor. *Science* **329**, 824 (2010).
17. M. Yi *et. al.*, Symmetry-breaking orbital anisotropy observed for detwinned $Ba(Fe_{1-x}Co_x)_2As_2$ above the spin density wave transition. *Proc. Natl. Acad. Sci. U.S.A.* **108**, 6878-6883 (2011).
18. M. Nakajima *et. al.*, Unprecedented anisotropic metallic state in undoped iron arsenide $BaFe_2As_2$ revealed by optical spectroscopy. *Proc. Natl. Acad. Sci. U.S.A.* **108**, 12238 (2011).
19. S. Jiang, H. S. Jeevan, J. Dong, P. Gegenwart, Thermopower as a Sensitive Probe of Electronic Nematicity in Iron Pnictides. *Phys. Rev. Lett.* **110**, 067001 (2013).
20. C. Xu, M. Müller, I. Sachdev, sing and spin orders in the iron-based superconductors. *Phys. Rev. B* **78**, 020501(R) (2008).
21. C. Fang *et al.*, Theory of electron nematic order in LaFeAsO. *Phys. Rev. B* **77**, 224509 (2008).
22. W. C. Lee, W. Lv, J. M. Tranquada, P. W. Phillips, Impact of dynamic orbital correlations on magnetic excitations in the normal state of iron-based superconductors. *Phys. Rev. B* **86**, 094516 (2012).
23. R. M. Fernandes, A. J. Millis, Nematicity as a Probe of Superconducting Pairing in Iron-Based Superconductors. *Phys. Rev. Lett.* **111**, 127001 (2013).
24. V. Oganesyan, S. A. Kivelson, E. Fradkin, Quantum theory of a nematic Fermi fluid. *Phys. Rev. B* **64**, 195109 (2001).
25. M. J. Lawler *et al.*, Nonperturbative behavior of the quantum phase transition to a nematic Fermi fluid. *Phys. Rev. B* **73**, 085101 (2006).
26. E. Fradkin *et al.*, Nematic Fermi Fluids in Condensed Matter Physics. *Annu. Rev. Condens. Matter Phys.* **1**, 153 (2010).
27. W. Bao, Physics picture from neutron scattering study on Fe-based Superconductors. *Chin. Phys. B* **22**, 087405 (2013).
28. Y. Liu *et al.*, Evolution of precipitate morphology during heat treatment and its implications for the superconductivity in $K_xFe_{1.6+y}Se_2$ single crystals. *Phys. Rev. B* **86**, 144507 (2012).
29. D. P. Shoemaker *et al.*, Phase relations in $K_xFe_{2-y}Se_2$ and the structure of superconducting $K_xFe_2Se_2$ via high-resolution synchrotron diffraction. *Phys. Rev. B* **86**, 184511 (2012).
30. X. Ding *et al.*, Influence of microstructure on superconductivity in $K_xFe_{2-y}Se_2$ and evidence for a new parent phase $K_2Fe_7Se_8$. *Nat. Commun.* **4**, 1897 (2013).

31. W. Li, *et al.,* Phase separation and magnetic order in K-doped iron selenide superconductor. *Nat. Phys.* **8**, 126-130 (2012).
32. Y. Texier *et al.,* NMR Study in the Iron-Selenide $Rb_{0.74}Fe_{1.6}Se_2$: Determination of the Superconducting Phase as Iron Vacancy-Free $Rb_{0.3}Fe_2Se_2$. *Phys. Rev. Lett.* **108**, 237002 (2012).
33. S. V. Carr *et al.,* Structure and composition of the superconducting phase in alkali iron selenide $K_yFe_{1.6+x}Se_2$. *Phys. Rev. B* **89**, 134509 (2014).
34. Z. Wang *et al.,* Archimedean solid-like superconducting framework in phase-separated $K_{0.8}Fe_{1.6+x}Se_2$ ($0 \leq x \leq 0.15$). http://arxiv.org/abs/1401.1001.
35. J. Zhao *et al.,* Neutron-Diffraction Measurements of an Antiferromagnetic Semiconducting Phase in the Vicinity of the High-Temperature Superconducting State of $K_xFe_{2-y}Se_2$. *Phys. Rev. Lett.* **109**, 267003 (2012).
36. T .K. Chen *et al.,* Fe-vacancy order and superconductivity in tetragonal $\beta$-$Fe_{1-x}Se$. *Proc. Natl. Acad. Sci. U.S.A.* **111**, 63-68 (2014).
37. T. Qian *et al.,* Absence of a Holelike Fermi Surface for the Iron-Based $K_{0.8}Fe_{1.7}Se2$ Superconductor Revealed by Angle-Resolved Photoemission Spectroscopy. *Phys. Rev. Lett.* **106**, 187001 (2011).
38. Y. Zhang *et al.,* Nodeless superconducting gap in $A_xFe_2Se_2$ (A=K, Cs) revealed by angle-resolved photoemission spectroscopy. *Nature Mater.* **10**, 273-277 (2011).
39. M. Yi *et al*., Observation of Temperature-Induced Crossover to an Orbital-Selective Mott Phase in $A_xFe_{2-y}Se_2$ (A=K, Rb) Superconductors. *Phys. Rev. Lett.* **110**, 067003 (2013)
40. A. M. Zhang *et al.,* Effect of iron content and potassium substitution in $A_{0.8}Fe_{1.6}Se_2$ (A=K, Rb, Tl) superconductors: A Raman scattering investigation. *Phys. Rev. B* **86**, 134502 (2012).
41. A. M. Zhang *et al.,* Two-magnon Raman scattering in $A_{0.8}Fe_{1.6}Se_2$ systems (A = K, Rb, Cs, and Tl): Competition between superconductivity and antiferromagnetic order. *Phys. Rev. B* **85**, 214508 (2012).
42. Y. J. Song *et al.,* Phase transition, superstructure and physical properties of $K_2Fe_4Se_5$. *Europhys. Lett.* **95**, 37007 (2011).
43. L. Sun *et al.,* Re-emerging superconductivity at 48 kelvin in iron Chalcogenides. *Nature* **483**, 67-69 (2012).
44. F. Chen *et al.,* Electronic identification of the parental phases and mesoscopic phase separation of $K_xFe_{2-y}Se_2$ superconductors. *Phys. Rev. X* **1**, 021020 (2011).

**Acknowledgments:**

We thank Dr. Ai-Hua Fang for fruitful discussion and valuable suggestions. This research cannot be done without the financial support from the Ministry of Science and Technology Grant NSC102-2112-M-001-025-MY3 and the Academia Sinica Thematic Research. Dr. Phillip M. Wu acknowledges support from AFOSR under DoDMURI Grant FA9550-09-1-0583.

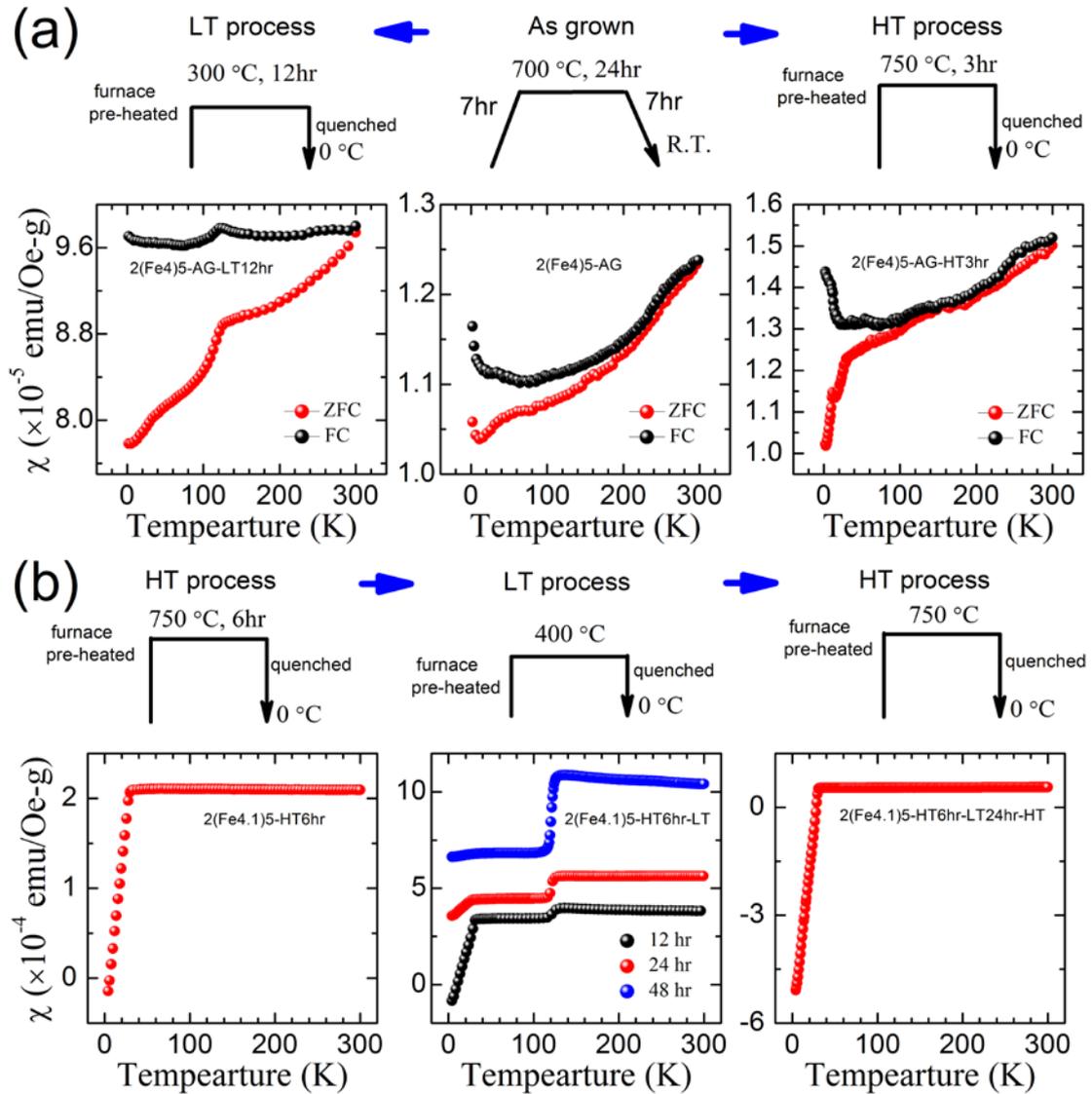

**Fig. 1.** (a) Temperature dependent magnetic susceptibility of polycrystalline $K_2Fe_4Se_5$ samples, with treatment procedure schematically shown. Middle panel is the result of theas grown sample. Left panel is for the (2(Fe4)5-LT) sample and the right panel is for the (2(Fe4)5-HT) sample. The sample quenched from 750 ºC shows two superconducting-like transitions at Tc = 31 K and 11 K. The sample annealed at 300 ºC exhibits a transition at 125 K. (b) Evolution of the magnetic susceptibility of 2(Fe4.1)5 sample. Left - The sample displays superconducting transition after HT treatment. Middle - The same sample can be converted to non-superconducting with the LT treatment, with more prominent125K transition with longer annealing time. Right – The sample becomes superconducting again, with increased volume fraction after another HT treatment.

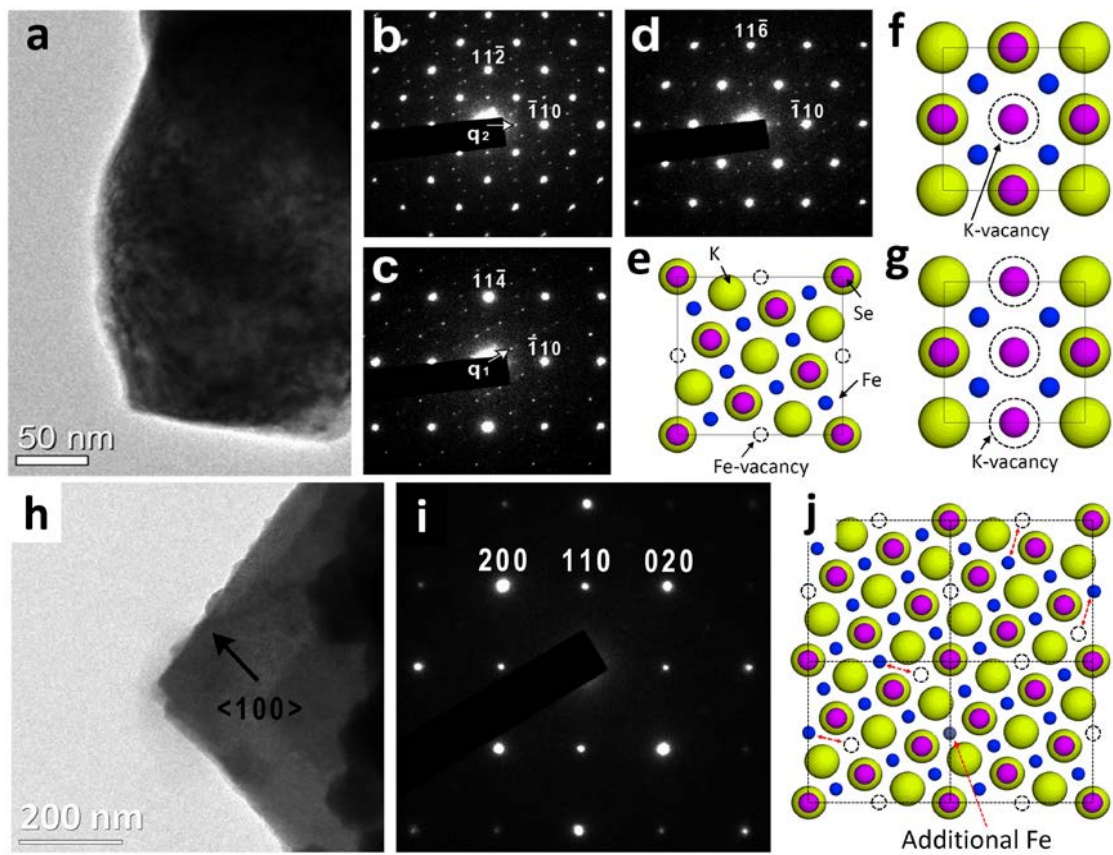

**Fig. 2**(a). TEM image of 2(Fe4)5-AG sample. Granular-type grains were found for this sample, see Fig. S7. SAED patterns were taken along (b) [111], (c) [221] and (d) [331] zone-axis directions. The c-axis SAED pattern that easily distinguishes Fe/K-vacancy orders is not available due to the limit of double-tilt holder. Yet two types of vacancy orders were still resolved as indicated: $q_1$ = 1/5(1,3,0), a typical $\sqrt{5} \times \sqrt{5}$ Fe-vacancy order in $K_2Fe_4Se_5$; $q_2$ = 1/2(1,1,0), a K-vacancy order in K-deficient $K_yFe_2Se_2$. (e-g) Structural models for $K_2Fe_4Se_5$, $K_{0.75}Fe_2Se_2$ and $K_{0.5}Fe_2Se_2$. (h) TEM image of $K_{1.9}Fe_{4.2}Se_5$ sample. Plate-like grains were found for this sample, see Fig. S7. (i) SAED pattern taken along [001] (c-axis) zone-axis direction. No vacancy order in either Fe or K types is observed. (j) Schematic diagram of a Fe-vacancy disorder configuration by changing few of the vacant and occupied Fe sites (4 pairs) and introducing additional Fe (one extra Fe).

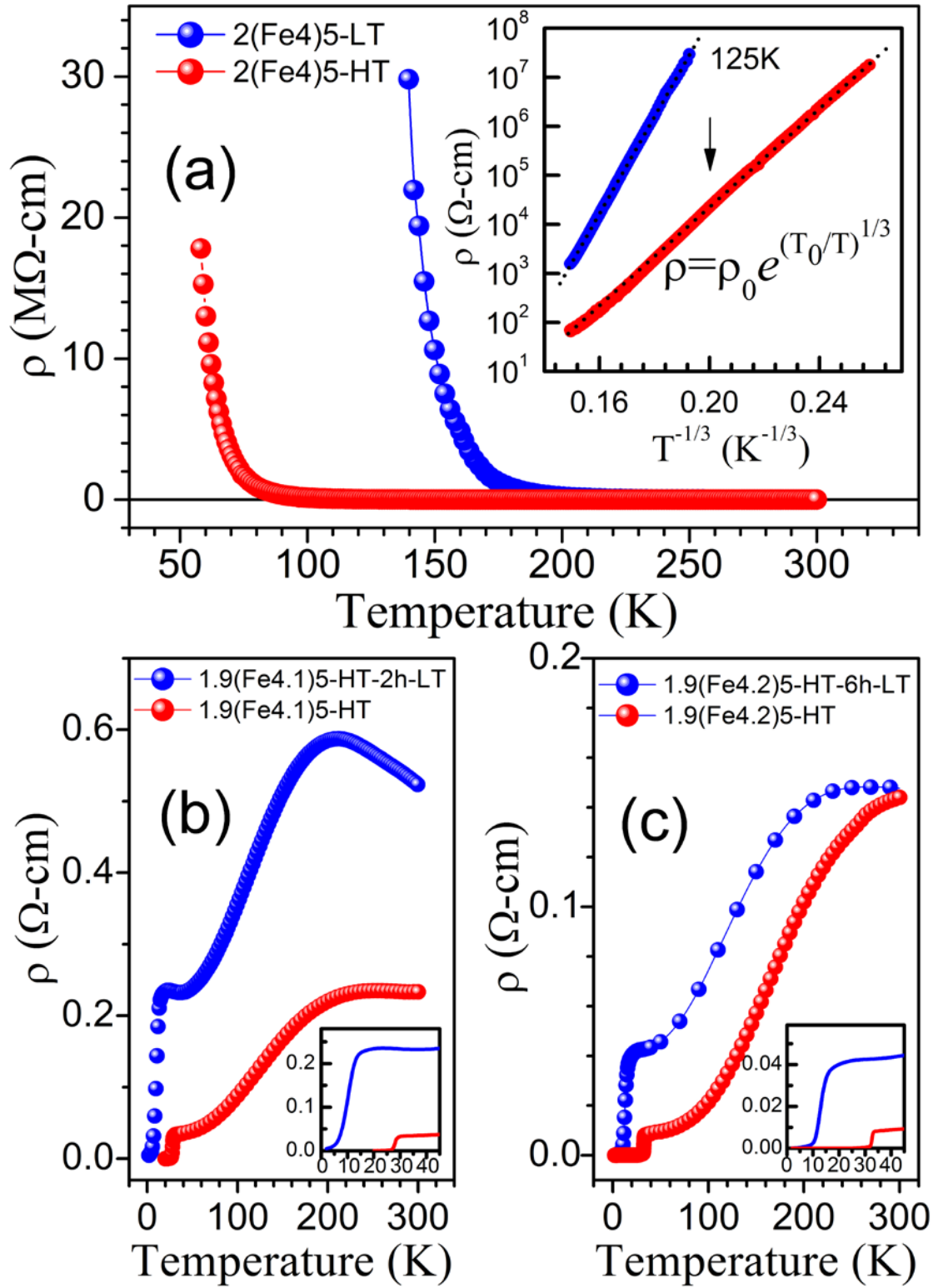

**Fig. 3.** Temperature dependent resistivity of $K_{2-x}Fe_{4+y}Se_5$. (a) The R-T curves of 2(Fe4)5-LT and 2(Fe4)5-HT samples show semiconducting behavior, which are fit to a 2-dimensional Mott variable-range-hopping model as shown in the inset. The "-LT" sample becomes extremely insulating close to the 125K transition observed in susceptibility, see Fig. 1(a). For the HT sample, which shows superconducting-like transitions in susceptibility measurement, the resistivity reduces substantially suggesting

the sample contains regions of superconducting puddles that are not connected. (b) R-T evolution of the 1.9(Fe4.1)5 samples. The HT procedure leads to superconducting sample with Tc ~ 31 K, while further LT annealing the sample reduces Tc and enhances the resistive peak, which along with the appearance of the 125K transition in magnetic measurement, see Fig. 1(b). (c) R-T evolution of the 1.9(Fe4.2)5 sample. The HT treated sample shows no resistive peak feature at high temperature. Further LT treatment reduces Tc and exhibits a resistive peak.

**Supplementary Materials:**

Materials and Methods

We establish a novel approach to combine the mechanical ball milling and high temperature calcination processes to prepare high quality stoichiometric $K_{2-x}Fe_{4+y}Se_5$ polycrystalline samples with homogenous chemical composition and crystal phase. We typically ball-mill the mixed powder of Fe, Se and K-plate with the exact stoichiometry for 60 minutes and repeated for 3 times. The well-mixed powder was sealed in quartz tube with argon and then put into furnace by slowly heated to 700°C in 7 hours, keep at the same temperature for 24 hours and then slow cooled to ambient temperature. We call these resulted powders the "As-Grown" samples. These As-grown samples were taken out of the quarts tube, regrind (with extreme care in the glove box to prevent from moisturized), sealed in quartz tube, and then further annealed in two different steps: one slowly heated to 300°C, kept in the furnace for 12 hours, then quenched in ice-water (the resulted materials are named –LT samples); the other step was to process the samples in a pre-heated (at 750°C) furnace directly with duration of 1~6 hours and followed by quenching in ice water (named –HT samples.) The nominal compositions we used in this study include $K_2Fe_4Se_5$ and $K_{1.9}Fe_{4.1}Se_5$ and $K_{1.9}Fe_{4.2}Se_5$. The samples we prepared are listed in table one together with the lattice parameters and superconducting temperatures.

In our experiment we observed the following conditions are the main reasons that formed the non-superconducting (Fe-vacancy order) phase. Conditions such as sample are not uniform, reaction temperature if below 750°C, reaction time is too long and furnace cooling rate could be an issue as well.

**Table S1** calculated structural parameters for $K_{2-x}Fe_{4+y}Se_5$ polycrystalline samples, obtained from X-ray powder diffraction data at room temperature.

| Sample | Conditions | Crystal Symmetry | $a$ (Å) | $c$ (Å) | $T_c$ (K) SQUID/RT | $T_M$ (K) |
|---|---|---|---|---|---|---|
| 2(Fe4)5-AG | Furnace cool-12h | I4/m | 8.7166(4) | 14.084(1) | – | – |
| 2(Fe4)5-LT | 300ºC-anneal-12 h | I4/m | 8.7202(4) | 14.078(1) | – | 125 |
| 2(Fe4)5-HT | 750ºC-quench-3 h | I4/m | 8.7102(6) | 14.099(1) | 29&11/– | – |
| 1.9(Fe4.1)5-HT | 750ºC-quench-6 h | I4/m | 8.7022(5) | 14.102(1) | 28.1/29.7 | – |
| 1.9(Fe4.2)5-HT-2h | 750ºC-quench-2 h | I4/mmm | 3.8809(5) | 14.110(5) | 30.8/30.9 | – |
| 1.9(Fe4.2)5-HT-6h | 750ºC-quench-6 h | I4/mmm | 3.8810(6) | 14.113(6) | 29.3/– | – |
| 1.9(Fe4.2)5-LT | 300ºC-anneal-16 h | I4/m | 8.688(2) | 14.097(5) | – | 125 |

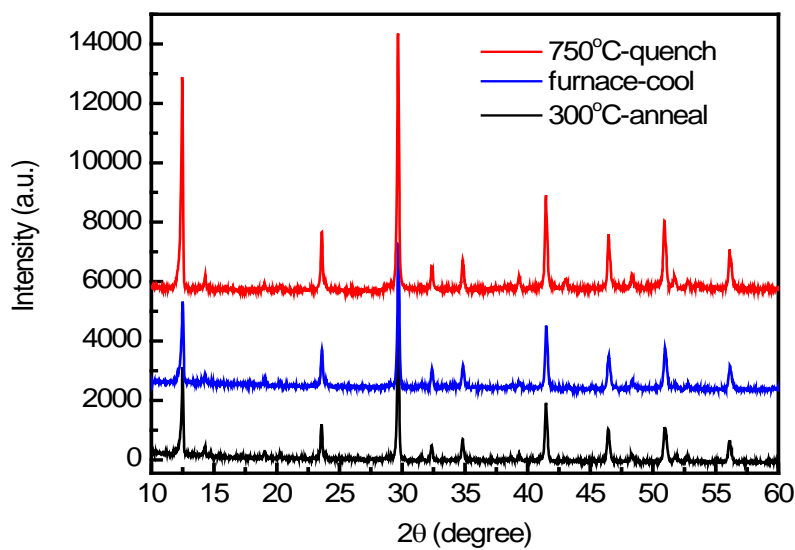

Figure S1. XRD of 2(Fe4)5 ($K_2Fe_4Se_5$) samples under different heat treatments: furnace-cooled from 750 ºC (blue), quenched from 750 ºC (red) and annealed at 300 ºC and then quenched to ice water (black). No significant difference among these patterns is observed, indicating no change in stochiometry.

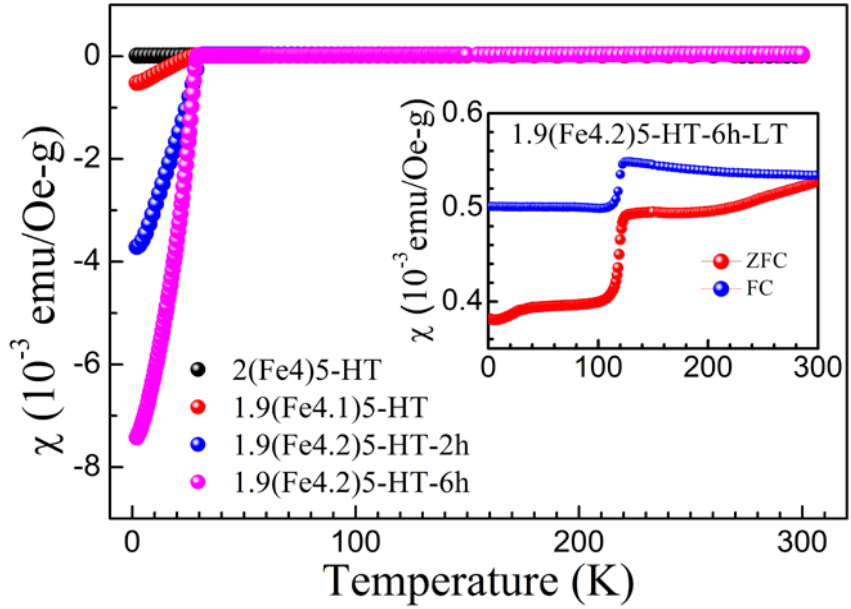

Figure S2. Temperature dependent magnetic susceptibility of HT-processed superconducting $K_{2-x}Fe_{4+y}Se_5$ samples. The superconducting volume ratio increases from 2(Fe4)5-HT sample to 1.9(Fe4.2)5-HT-6h sample. Inset shows the suppression of superconductivity for 1.9(Fe4.2)5-HT-6h sample that was further post-annealed at 300 ºC for 16 hours (1.9(Fe4.2)5-HT-6h-LT). The magnetic behavior is very much like that of a 2(Fe4)5-LT sample shown in Fig. 1, with a transition at ~ 125 K.

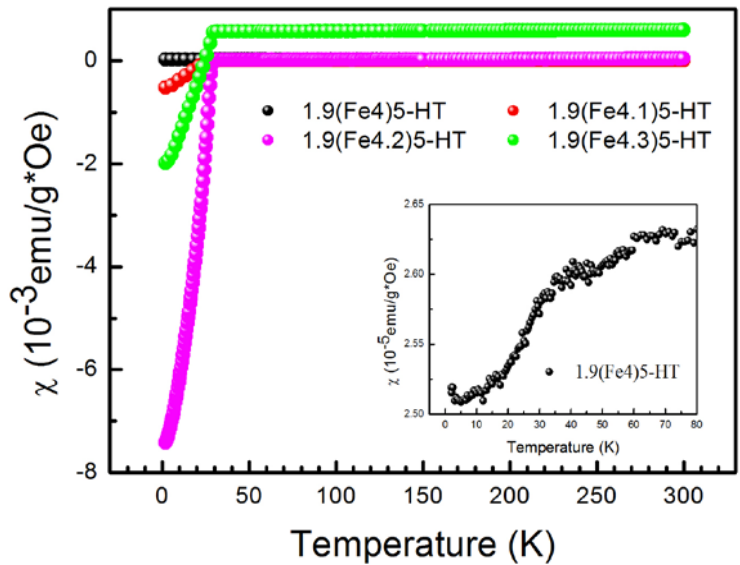

Figure S3. Magnetic susceptibility for a series of $K_{2-x}Fe_{4+y}Se_5$. The superconducting volume fraction increases as extra Fe is added to 4.2. However, further increase Fe results in reducing superconducting volume, as the green curve for 1.9(Fe4.3)5-HT shown. The inset shows detail of the superconducting transition of the 1.9(Fe4)5-HT

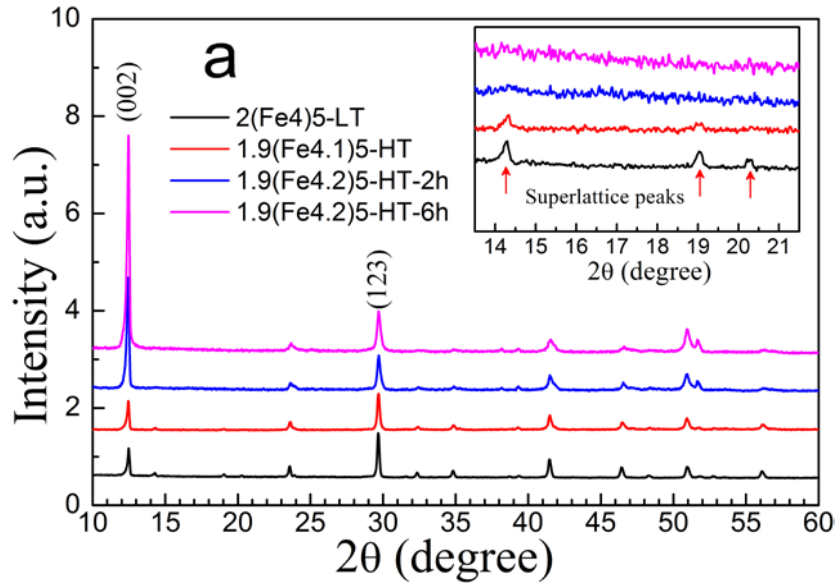

Figure S4. XRD patterns of polycrystalline $K_{2-x}Fe_{4+y}Se_5$ samples, revealing no phase separation within the instrument resolution. Inset enlarges 2θ in between 13º and 22º where superstructure peaks from $\sqrt{5} \times \sqrt{5}$ Fe-vacancy order are most profound. The suppression of the superstructure and Fe-vacancy order is accompanied by a monotonic decrease and increase for peak (123) and peak (002), respectively.

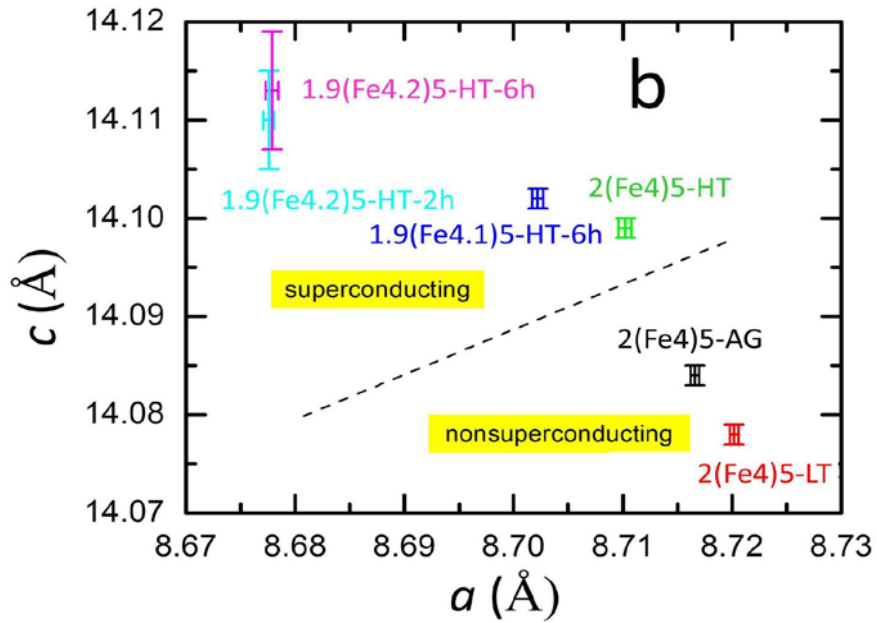

Figure S5. Lattice parameters for polycrystalline $K_{2-x}Fe_{4+y}Se_5$ samples with different preparation processes.

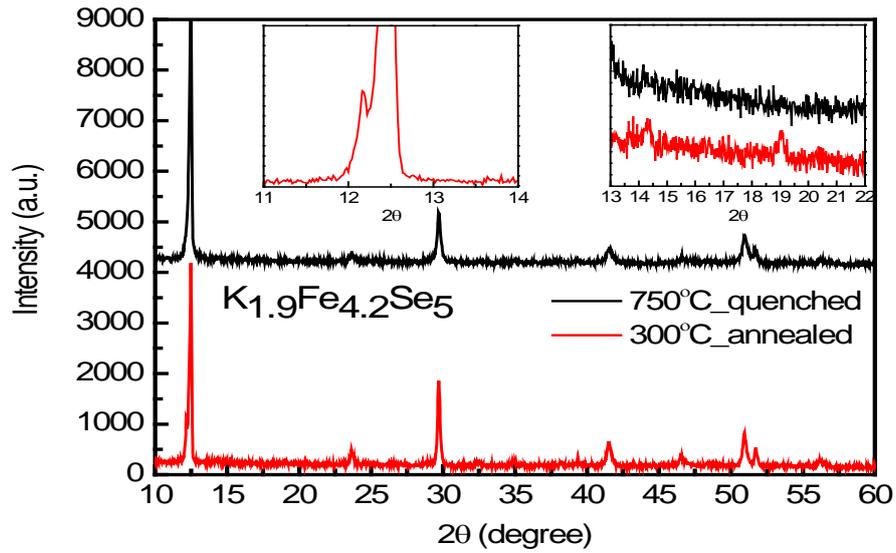

Figure S6. XRD patterns of polycrystalline $K_{1.9}Fe_{4.2}Se_5$ samples. Black curve shows the sample quenched from 750 ºC, which is superconducting. Red curve shows the sample post-annealed at 300 ºC and then quenched to ice water, which is not superconducting. Left inset shows a closer look of 2θ in the range of 11º–14º, indicating a minority second phase with expanded c-axis. Right inset zoom in between 13º and 22º of 2θ, revealing the emergence of superstructure peaks after annealing at 300 ºC.

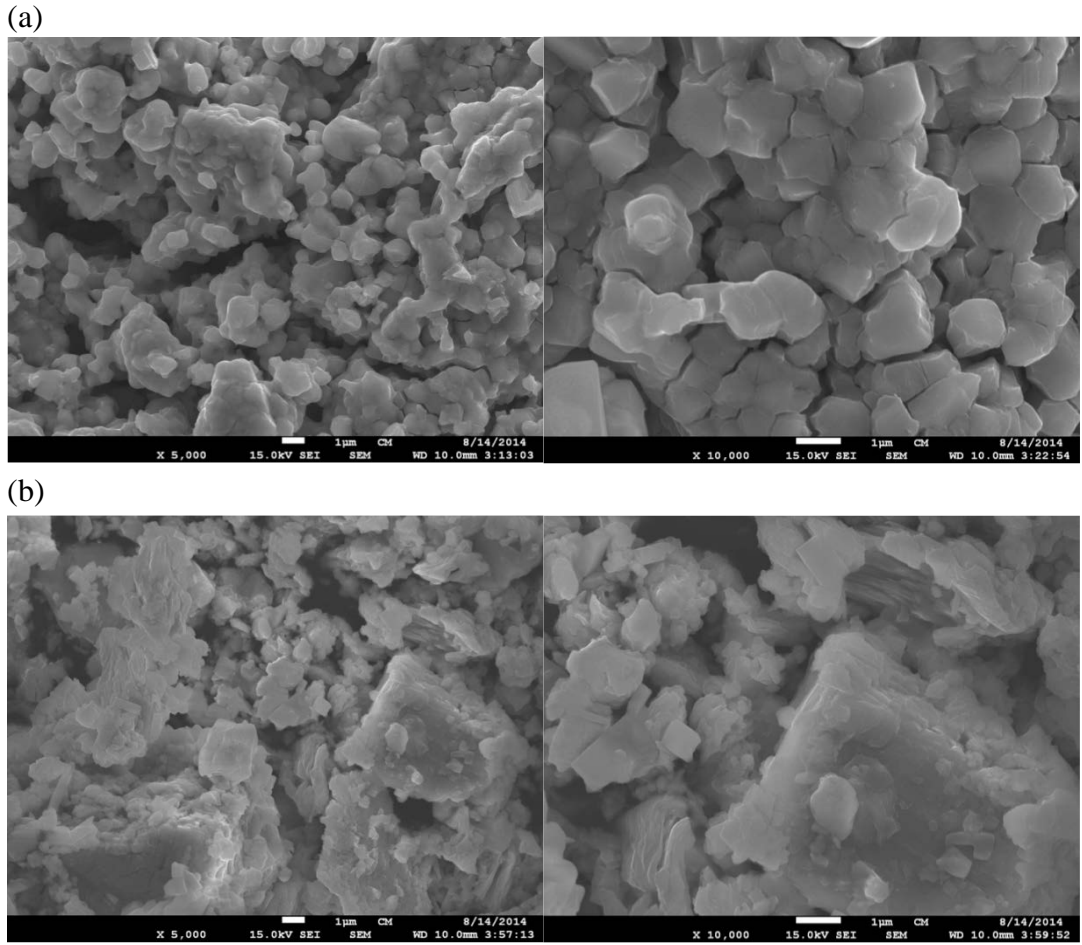

Figure S7. SEM images of (a) furnace-cooled $K_2Fe_4Se_5$ and (b) $K_{1.9}Fe_{4.2}Se_5$ sample. The granular crystal morphology for the former and layered feature for the latter is evident.

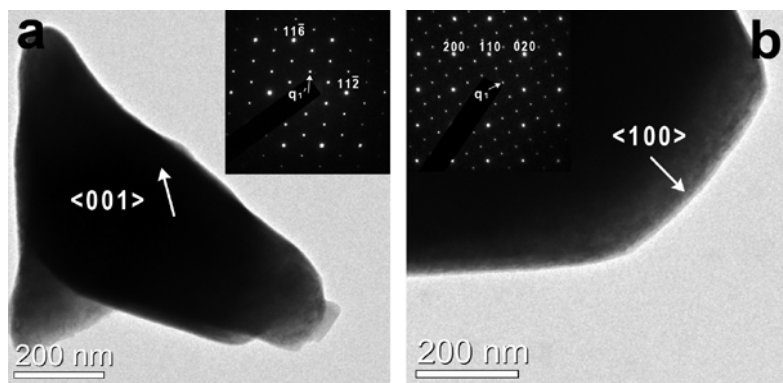

Figure S8. Two TEM images of the 2(Fe4)5-LT samples. The grain morphology is basically the same as the 2(Fe4)5-AG samples. (a) A grain away from c-axis. Inset in (a) shows the SAED pattern taken along [-42-1] zone-axis direction. Strong superstructure spots $q_1' = 1/5(1,3,10)$ corresponds to the $\sqrt{5} \times \sqrt{5}$ Fe-vacancy order in $K_2Fe_4Se_5$. (b) A grain along c-axis zone pattern. Clear single domain $\sqrt{5} \times \sqrt{5}$ Fe-vacancy order with modulation of $q_1 = 1/5(1,3,0)$ is observed in the SAED pattern, as shown in the inset. No K-vacancy order is observed in this sample.